\documentclass[twocolumn,aps,prl,floats]{revtex4}
\usepackage{graphicx}
\usepackage{amssymb}
\usepackage{epstopdf}
\usepackage{epsfig}
\begin{document}

\title{Should you believe that this coin is fair?}

\author{William Bialek}

\affiliation{Joseph Henry Laboratories of Physics and the Lewis--Sigler Institute for Integrative Genomics\\
Princeton University, Princeton, New Jersey 08544}

\begin{abstract}
Faced with a sequence of $N$ binary events, such as coin flips (or Ising spins), it is natural to ask whether these events reflect some underlying dynamic signals or are just random.   Plausible models for the dynamics of hidden biases lead to surprisingly high probabilities of misidentifying random sequences as biased.  In particular, this probability decays as $N^{-1/4}$, so that no reasonable amount of data would be sufficient to induce the concept of a fair coin with high probability.  I suggest that these theoretical results may be relevant to understanding experiments on the apparent misperception of random sequences by human observers.
\end{abstract}

\date{\today}

\maketitle

There is a large literature testifying to the errors that humans make in reasoning about probability \cite{heuristics+biases, reviews}.  Perhaps most fundamental is the claim that people routinely detect order and hidden causes in genuinely random sequences \cite{hot_hand}.   These apparent limitations on human rationality have broad implications, not least for economics, and have attracted considerable attention in the popular press.   In contrast with these results, a number of experiments indicate that humans and other animals can change their behavior in response to changes in the probabilities of stimuli and rewards, sometimes making optimal use of the available data 
\cite{criteria,carpenter+williams_95,gallistel+al_01,sugrue+al_04}.  Similarly, many perceptual discriminations approach the limits to reliability set by noise near the sensory input \cite{barlow_81,bialek_87,bialek_02},  and related ideas of statistical optimization have emerged in recent work on motor control \cite{harris+wolpert_88,kording+wolpert_04,trommerhauser+al_05}.  There is even the suggestion that if  the detection of order vs. randomness is cast in the standard two--alternative format for perceptual discrimination experiments, then people can learn to perform with close to the statistically maximum reliability \cite{lopes+oden_87}.
While  nearly perfect neural processing of statistical data under some conditions could coexist with qualitative failures at similar problems under different conditions,
it would be attractive to have a more unified view of the brain  as an engine for probabilistic inference.

The problem of identifying genuinely random sequences has a number of subtleties that seem not to have been emphasized in the previous literature.
In particular, from a Bayesian point of view our confidence that a given sequence really was generated at random depends entirely on the universe of alternative models that we are willing to consider.  Here I consider a family of models that involve time--dependent biases of unknown magnitude, similar in spirit to the models that would allow optimal inference under the conditions of the experiments in Refs \cite{gallistel+al_01,sugrue+al_04}, and show that an observer who tries to understand the world using these models will exhibit a surprisingly large probability of misidentifying random sequences as having small but nonzero biases.
Most importantly this probability declines only as a fractional power of the sequence length, so that (for example) no reasonable human experience with the flipping of a coin would provide sufficient evidence to induce the concept of fairness with high probability.

To be concrete let us consider data in the form of binary sequences---coin flips (heads/tails), for example, or a sequence of rewards/nonrewards from a particular class of actions.  Let the observations come in a sequence labeled by ${\rm n} = 1, 2, \cdots , N$, and let the binary variable on each observation $\rm n$ be denoted by $\sigma_{\rm n}$.  If the process is described by a fair coin then all sequences occur with equal probability,
\begin{equation}
P(\{\sigma_{\rm n}\}|{\rm fair\ coin}) = {1\over{2^N}} .
\label{def_fair}
\end{equation}
The problem faced by the observer, however, is to estimate the probability that this particular sequence was generated by a fair coin, that is $P({\rm fair\ coin}|\{\sigma_{\rm n}\})$.  Bayes' rule tells us that we can write this probability as
\begin{equation}
P({\rm fair\ coin}|\{\sigma_{\rm n}\}) = {{P(\{\sigma_{\rm n}\}|{\rm fair\ coin}) P({\rm fair\ coin})}\over{P(\{\sigma_{\rm n}\})}} ,
\end{equation}
where $P({\rm fair\ coin})$ measures the a priori probability of a fair coin and $P(\{\sigma_{\rm n}\})$ measures the probability that this particular sequence will arise, averaged over all possible models that might describe it; schematically we can write
\begin{equation}
P(\{\sigma_{\rm n}\}) = \sum_{\rm all\ models} P(\{\sigma_{\rm n}\}|{\rm model}) P({\rm model}) ,
\end{equation}
and of course one of the possible models is the fair coin.  Because the fair coin generates all sequences with equal probability [Eq (\ref{def_fair})], the probability that a fair coin generated a given sequence depends on the details of the sequence only through the term $P(\{\sigma_{\rm n}\})$, {\em i.e.} only through the average over all possible models of the sequence.  Thus our confidence that our experience is described by a fair coin depends entirely on the set of alternative models that we think is appropriate to the situation.

A plausible set of alternatives to the fair coin is that each $\sigma_{\rm n}$ responds independently to a bias,  but this bias may change over time.  To specify this class of models completely requires at least three steps.  First we need to describe the bias on each trial. To treat heads and tails symmetrically we can represent heads/tails as an Ising spin $\sigma_{\rm n} = \pm 1$, and measure the bias on observation $\rm n$ as an effective magnetic field $h_{\rm n}$ such that \cite{Ising}
\begin{equation}
P(\sigma_{\rm n}|h_{\rm n})  = {{\exp(-h_{\rm n}\sigma_{\rm n})}\over{2\cosh(h_{\rm n})}}.
\end{equation}
Assuming that the bias acts on each observation independently, we have
\begin{eqnarray}
P(\{\sigma_{\rm n}\} | \{h_{\rm n}\}) &=& \prod_{{\rm n}=1}^N
{{\exp(h_{\rm n}\sigma_{\rm n})}\over{2\cosh(h_{\rm n})}} = {1\over{2^N}} {\rm e}^{-f(\{\sigma_{\rm n} ;
h_{\rm n}\})}
\\
f(\{\sigma_{\rm n} ;
h_{\rm n}\}) &=& 
\sum_{{\rm n}=1}^N \left[ \ln\cosh(h_{\rm n}) +  h_{\rm n}\sigma_{\rm n}  \right] . 
\end{eqnarray}
A particular set of biases $\{h_{\rm n}\}$ constitutes one possible model.  The second step is that to average over models we will need a hypothesis about the distribution of these biases.  Consider the case where the bias is on average zero (coins or circumstances that favor heads are as likely as those that favor tails), and where the presumably small fluctuations in bias are drawn from a Gaussian with root--mean--square fluctuations $h_{\rm rms}$. Dynamics of the bias vs.~time are described by a correlation function $C(\tau )$,
$
\langle h_{\rm n} h_{\rm m} \rangle = h_{\rm rms}^2 C({\rm n} - {\rm m}).
$
If, for example, the bias tends to stay constant for runs of $10$ observations, then $C(\tau)$ should be close to $1$ for $|\tau | \leq 10$ and fall to zero for $|\tau | \gg 10$.
It is convenient to the think about a matrix $\hat C$ with elements defined by $(\hat C)_{\rm nm} = C({\rm n} - {\rm m})$.
Then the full distribution of biases has the form
\begin{eqnarray}
P(\{h_{\rm n}\}|h_{\rm rms}) &=&{1\over Z}
\exp\left[
-{1\over{2h_{\rm rms}^2}} \sum_{{\rm n, m}=1}^N h_{\rm n} (\hat C^{-1})_{\rm nm} h_{\rm m}
\right] ,\nonumber\\
Z &=& (\sqrt{2\pi} h_{\rm rms})^{N} \exp\left[ 
+{1\over 2} {\rm Tr}\,\ln \hat C\right] .
\end{eqnarray}
As a third and final step we have to specify our knowledge of $C$ and $h_{\rm rms}$.  To keep things simple let us assume that there is something about the situation which makes us certain about the time scales on which bias can vary, so that we know the correlation function $C(\tau )$, but we don't know for sure how strong the biases can get, so we have to average over some distribution $P(h_{\rm rms})$.

Putting the various terms together, we can write the probability of observing a sequence $\{\sigma_{\rm n}\}$ in the broad family of biased models as
\begin{eqnarray}
&&P(\{\sigma_{\rm n}\} |{\rm biased}) \nonumber\\
&&\,\,\,\,\, =\int dh_{\rm rms} d^N h_{\rm i} \,P(\{\sigma_{\rm n}\}|\{h_{\rm n}\}) P(\{h_{\rm n}\}|h_{\rm rms}) P(h_{\rm rms}), \nonumber\\
&&
\label{integral}
\end{eqnarray}
and then
\begin{eqnarray}
P(\{\sigma_{\rm n}\}) &=& P(\{\sigma_{\rm n}\} |{\rm fair\ coin}) P({\rm fair\ coin})\nonumber\\
&&\,\,\,\,\,\,\,\,\,\,
+
P(\{\sigma_{\rm n}\} |{\rm biased}) P({\rm biased}).
\end{eqnarray}
Thus the probability of an observed sequence arising from a fair coin is
\begin{equation}
P({\rm fair\ coin}|\{\sigma_{\rm n}\}) = {1\over{1 + \exp(\Lambda - \mu)}},
\label{lambda-mu}
\end{equation}
where the log--likelihood ratio is
\begin{equation}
\Lambda = \ln[P(\{\sigma_{\rm n}\} |{\rm biased})/P(\{\sigma_{\rm n}\} |{\rm fair\ coin})] ,
\end{equation}
and the threshold $\mu = \ln[P({\rm fair\ coin})/P({\rm biased})]$.

In the limit that $h_{\rm rms}$ is small we can compute the integral over $\{h_{\rm n}\}$ in Eq (\ref{integral}) as a perturbation series. To fourth order in $h_{\rm rms}$, the result is
\begin{eqnarray}
&&\hskip - 0.3 in \int d^N h_{\rm i} \,P(\{\sigma_{\rm n}\}|\{h_{\rm n}\}) P(\{h_{\rm n}\}|h_{\rm rms}) \nonumber\\
&&\,\,\,\,\,\,\,\,\,\,
={1\over{2^N}}\exp\left[-{{N\tau_c}\over 2}a h_{\rm rms}^4 
+ \sqrt{N\tau_c} z  h_{\rm rms}^2  \right] ,
\label{exponent}
\end{eqnarray}
where the correlation time is defined by
\begin{equation}
\tau_c = {1\over{2N}}{\rm Tr} \,\hat C^2 = {1\over 2}\sum_{\rm n} C^2({\rm n}),
\end{equation}
$a = 1 +y \sqrt{4g/N\tau_c}$, $g = ({\rm Tr} \,\hat C^4)/({\rm Tr} \,\hat C^2)$, and the variables $z$ and $y$ depend on the particular sequence that we observe: \begin{eqnarray}
z &=& {1\over\sqrt{4N\tau_c}}\left[\sum_{\rm nm} \sigma_{\rm n} \hat C_{\rm nm} \sigma_{\rm m} - N\right]
\label{defz}\\
y &=& {1\over\sqrt{4gN\tau_c }} \left[\sum_{\rm nm} \sigma_{\rm n} (\hat C^2)_{\rm nm} \sigma_{\rm m} - {\rm Tr}\, \hat C^2\right] .
\nonumber\\
&&
\end{eqnarray}
The normalization of $z$ and $y$ is chosen so that they each have zero mean and unit variance if the $\{\sigma_{\rm n}\}$ actually are generated by a fair coin.
The exponential in Eq (\ref{exponent}) sets a scale for $h_{\rm rms} \sim N^{-1/4}$, which becomes small at large $N$.  Thus when we do the integral over $h_{\rm rms}$ in Eq (\ref{integral}) it will be dominated by very small values of $h_{\rm rms}$, so the relevant parameter is $P(h_{\rm rms} \rightarrow 0) = \rho$ \cite{rho=0?}.  Then
\begin{eqnarray}
&&\hskip - 0.4 in
P(\{\sigma_{\rm n}\} |{\rm biased})  \nonumber\\
&&\hskip -0.25 in ={\rho\over{2^N}}
\int dh_{\rm rms} \exp\left[-{{N\tau_c}\over 2}a h_{\rm rms}^4 
+ \sqrt{N\tau_c} z  h_{\rm rms}^2  \right] .
\end{eqnarray}
At large $N$ and $z<0$ the integral is dominated by its behavior near $h_{\rm rms}=0$, while for $z>0$ it is dominated by a saddle point at
$h_* = (z^2/a^2N\tau_c)^{1/4}$.  When the dust settles, the large $N$ approximation to the log--likelihood ratio becomes
\begin{eqnarray}
\Lambda (z< 0) &\sim& {1\over 4}\ln\left[{{\pi^2\rho^4}\over{4z^2 N\tau_c} }\right] \\
\Lambda (z>0) &\sim& {{z^2}\over{2a}} + {1\over 4}\ln\left[{{\pi^2\rho^4}\over{4z^2 N\tau_c}}\right] ,
\end{eqnarray}
where the dependence on the observed sequence is through the value of $z$ (and, negligibly in this limit, also through the dependence of $a$ on $y$).

From Eq (\ref{lambda-mu}) we see that if $\Lambda > \mu$ then the probability of the sequence $\{\sigma_{\rm n}\}$ being described by a fair coin is less than half.  Equivalently, if $\Lambda > \mu$ then the data are more likely to be described by a biased model.  If we want to make correct assignments with highest probability, then the correct decision rule is maximum likelihood \cite{criteria}, so that all sequences with $\Lambda > \mu$ should be labeled as biased.  Obviously there is some probability that this condition is met even if the sequence in fact was generated at random.  Note that for large $N$ the variable $z$, which determines $\Lambda$, is the sum of  many terms which are independent if the sequence really is random.  Thus the distribution of $z$ approaches a Gaussian (with zero mean and unit variance, by construction) and we can estimate the probability that $\Lambda > \mu$, which is the probability that a random sequence will be identified as biased \cite{careful_w/gauss}.

The condition $\Lambda > \mu$ corresponds either to
$z- < z < 0$ or $z > z_+$, where
\begin{eqnarray}
z_- &=& - {\pi \over 2} \left( \rho {{P({\rm fair\ coin})}\over{P({\rm biased})}} \right)^2 {1\over\sqrt{N\tau_c}} \\
z_+ &=& \ln^{1/2}\left[ {{2 z_+ \sqrt{N\tau_c}}\over{\pi \rho^2}}\left(  {{P({\rm fair\ coin})}\over{P({\rm biased})}}\right)^2\right]
\end{eqnarray}
Then a sequence generated by a fair coin will be assigned as biased with a probability given by 
\begin{equation}
P_{\rm error} \approx \int_{-z_-}^0 {{dz}\over{\sqrt{2\pi}}}
 {\rm e}^{-z^2 / 2}
+ \int_{z_+}^\infty  {{dz}\over{\sqrt{2\pi}}} {\rm e}^{-z^2 / 2} .
\end{equation}
For large $N$, $|z_-| \ll 1$ and $z_+ \gg 1$, which lets us approximate the integrals to find
\begin{equation}
P_{\rm error}
\sim r^2 \sqrt{\pi\over{8N\tau_c}}
+ {r\over 2} \cdot {1\over{(N\tau_c )^{1/4}}} \cdot {1\over{\ln^{5/8}\left[ 2 \sqrt{N\tau_c}/\pi r^2\right]}} ,
\end{equation}
where $r = \rho P({\rm biased})/P({\rm fair\ coin})$ and I neglect terms of the form $\ln\ln N$.  At sufficiently large $N$ the second term dominates, and we have simply
\begin{equation}
P_{\rm error}
\sim  \left( {{N_c}\over N}\right)^{1/4} \cdot {1\over{\ln^{5/8}\left[ \sqrt{N/(4\pi^2 N_c)}\right]}} ,
\label{result}
\end{equation}
where the scale is set by $N_c = r^4/(16\tau_c)$.

The power--law decay of the error probability is surprising if we have in mind the conventional problem of statistical hypothesis testing.  With two specific alternative hypotheses, e.g. the fair coin and a coin with fixed (and known) bias, the log--likelihood ratio on average grows linearly with the number of examples that we observe, this growth rate being the Kullback--Leibler divergence between the two hypothesized distributions, and the resulting error probability falls exponentially.  By allowing for the possibility that biases change with time over the course of our observations---which seems plausible in many natural contexts---we construct a family of models with a number of parameters that grows as the size of the data set increases.    In addition, the particular model of bias considered here allows larger data sets to be used to test for weaker biases (smaller $h_{\rm rms}$), which again is plausible but very different from the simpler problem of discriminating between two fixed distributions.  To summarize, if we test data for a fixed bias of known size, then the probability of mistakenly finding order in a random sequence decays exponentially.  If on the other hand we test for time dependent biases of unknown magnitude, the error probability falls as a power--law.  The case considered here generates a $1/4$ power, but one can construct models that generate other powers as well \cite{rho=0?}.

In the case of two specific hypotheses, the exponential dependence of error probability also means that the trading between number of examples and prior probability is only logarithmic.  Suppose that when fair and biased coins are equally likely a priori ($\mu =0$), it takes $N_0$ examples to reach some criterion level of error.  If we now imagine that truly fair coins are rare, in the conventional hypothesis testing view it will take $\Delta N_0 \propto \ln[P({\rm biased})/P({\rm fair\ coin})]$ additional examples to reach the same level of error.  In contrast, when we search for time dependent biases with unknown magnitude, we are much more sensitive to prior probabilities, since Eq (\ref{result}) predicts $N_0 \propto N_c$, or $N_0 \rightarrow N_0 [P({\rm biased})/P({\rm fair\ coin})]^4$.  Concretely, if we can expect that genuinely random sequences constitute only $\sim 10\%$ of the events that we will see, that the typical scale of biases is $h_{\rm rms} \sim 1$ so that $\rho \sim 1$, and that correlation times are $\tau_c \sim 10$ events, then $N_c \sim 40$ and hence even at $N=10^4$ the error probability is $P_{\rm error}\sim 0.25$ \cite{value}.

The sequences which are identified as biased are primarily those with large positive values of $z$ from Eq (\ref{defz}).  If the expected correlation function of the bias is everywhere positive (so that we don't expect oscillating biases), then this singles out sequences that have an excess of runs with multiple heads or tails in a row.  Put another way, sequences which are declared to be random have fewer runs than expected in genuinely random distributions.  This is in agreement with experiments showing that when human subjects are asked to generate random sequences or to assess sequences for randomness, they behave according to a ``representativeness heuristic'' \cite{KT_72} or as if there were a ``law of small numbers'' \cite{TK_71} according to which short sequences are more typical of long sequence mean behavior than actually predicted for a fair coin.

All of the analysis done here can be repeated in a model where ``bias'' is replaced with serial correlation.  Then the natural parameter is not a local magnetic field but a local exchange interaction $J_{\rm n}$ between neighboring spins $\sigma_{\rm n}$ and $\sigma_{\rm n-1}$.  The analog of $z$ [Eq (\ref{defz})] which controls the likelihood of a sequence being assigned as random or correlated is played by
\begin{equation}
z' = {1\over\sqrt{4N\tau_c}} \left[\sum_{\rm nm}( \sigma_{\rm n} \sigma_{\rm n+1}) {\hat C}_{\rm nm} (\sigma_{\rm m} \sigma_{\rm m+1}) - N\right] ,
\end{equation}
where $\hat C$ is now the correlation matrix and $\tau_c$ is the correlation time for fluctuations in $J$.  Again the sequences identified as biased will be those with large positive $z'$, corresponding to an excess of {\em either} repetitions or alternations.  This may provide an even more accurate description of human perceptual biases toward representativeness in small samples \cite{reviews, KT_72}.

The model of dynamic biases considered here is closely related to the experiments in Refs \cite{gallistel+al_01} and \cite{sugrue+al_04}, where animals experience time dependent biases in reward probability and have to modulate their behaviors accordingly.  In these experiments the possibility of a fair coin is excluded by construction, so the question is not to determine whether the bias exists but rather to determine its current value and use this value in decision making \cite{sort_of}.  Within the model above one can show that, for  weak biases, the optimal estimate is
$
{\hat h}_{\rm n} = \sum_{\rm i \geq 0} K({\rm i}) \sigma_{\rm n-i} ,
$
where the kernel $K$ is determined by
\begin{equation}
\sum_{\rm m} {\hat C}_{\rm nm} K({\rm m-j}) + {1\over{h_{\rm rms}^2}} K({\rm n-j}) = {\hat C}_{\rm nj}
\label{kernel}
\end{equation}
Specifically, if $C(\tau ) = \exp(-|\tau|/\tau_c)$, then the kernel is also exponential,
$K(\tau ) \propto \exp(-\tau/\tau_{\rm int})$, where the integration time is
$\tau_{\rm int} = \tau_c/\sqrt{1 + 2h_{\rm rms}^2\tau_c}$.  Thus, when correlation times are long,  even if $h_{\rm rms}$ is not too large the optimal estimator integrates for a time much shorter than the correlation time---e.g., with correlation times over order $100$ events,  $\tau_{\rm int} \sim 10-15$ events,  in reasonable agreement with the results of Ref \cite{sugrue+al_04}.

In the framework considered here, the failure to recognize genuinely random sequences arises not as a limitation but rather as an inevitable consequence of the optimal search for weak dynamic biases.  Important tests of this idea thus include generalizations of the experiments in Refs \cite{gallistel+al_01,sugrue+al_04} that allow detailed comparisons of human strategies for bias estimation with the optimal strategy, especially the prediction that the optimal strategy shifts with the context defined by the distribution out of which the fluctuating biases are drawn.
Although not a complete theory, it seems plausible that the objective difficulty in inducing the concept of a truly fair coin found here could be related to other problems in reasoning about probability that usually are ascribed to subjective factors.

\acknowledgements{I thank GS Corrado, A Montagnini, R Rumiati, RR de Ruyter van Steveninck, SP Strong, LP Sugrue  and WT Newsome for helpful discussions.  Recent arguments with A Elga and D Osherson provided a stimulus to finish up; thanks to the Princeton University Council on the Humanities for the events that catalyzed these arguments. This  work was supported in part by  National Science Foundation Grant IIS--0423039, through  the program for Collaborative Research in Computational Neuroscience.}


\begin{thebibliography}{99}
\bibitem{heuristics+biases}
D Kahneman, P Slovic \& A Tversky, eds. {\em Judgement Under Uncertainty:  Heuristics and Biases} (Cambridge University Press, Cambridge, 1982).
\bibitem{reviews}
MA Bar--Hillel \& WA Wagenaar, in {\em A Handbook for Data Analysis in the Behavioral Sciences: Methodological Issues,} G Keren \& C Lewis, eds., pp. 369--393 (Erlbaum, Hillsdale NJ, (1993); RS Nickerson, {\em Psych Rev} {\bf 109,} 330--357 (2002).
\bibitem{hot_hand}
A classic example is the widespread belief that basketball players can have a ``hot hand;''  see T Gilovich, R Vallone \& A Tversky, {\em Cognitive Psych} {\bf 17,} 295--314 (1985).
\bibitem{criteria}
DM Green \& JA Swets,  {\em Signal Detection Theory and Psychophysics}   (Wiley, New York, 1966).
\bibitem{carpenter+williams_95}
RHS Carpenter  \& MLL Williams,    {\em Nature} {\bf 377}, 59--62 (1995).
\bibitem{gallistel+al_01}
CR Gallistel, TA Mark, AP King \& PE Latham,  Ê  {\em J Exp Psych: Animal Behav} {\bf Ê27,} 354--372 (2001).
\bibitem{sugrue+al_04}
LP Sugrue, GS Corrado \& WT Newsome,    {\em Science} {\bf 304,} 1782--1787 (2004).
\bibitem{barlow_81}
HB Barlow,  {\em Proc R Soc Lond Ser B} {\bf  212,} 1--34 (1981).
\bibitem{bialek_87}
W Bialek,  {\em Ann Rev Biophys Biophys Chem} {\bf 16,} 455--478 (1987).
\bibitem{bialek_02}
W Bialek, in  {\em Physics of Biomolecules
and Cells: Les Houches Session LXXV,} H Flyvbjerg, F J\"ulicher, P
Ormos \& F David, eds, pp 485--577 (EDP Sciences, Les Ulis;
Springer--Verlag, Berlin, 2002).  Also available at http://arXiv.org/abs/physics/0205030.
\bibitem{harris+wolpert_88}
CM Harris \& DM Wolpert,  {\em Nature} {\bf 394,} 780--784 (1988).
\bibitem{kording+wolpert_04}
KP Kording \& DM Wolpert, {\em Nature} {\bf 427,} 244--247 (2004).
\bibitem{trommerhauser+al_05}
J Tormmerha\"user, S Gephstein, LT Maloney, MS Landy \& MS Banks, {\em J Neurosci} {\bf 25,} 7169--7178 (2005).
\bibitem{lopes+oden_87}
LL Lopes \& GC Oden,  {\em J Exp Psych: Learning, Memory, and Cognition} {\bf 13}, 392--400 (1987).
\bibitem{Ising}
Although this is a natural formulation for physicists, it is worth emphasizing  any bias parameter can always be rewritten as an effective magnetic field in this notation.  
\bibitem{rho=0?}
If $P(h_{\rm rms})$ vanishes as $\sim h_{\rm rms}^\alpha$ (or diverges for $\alpha < 0$), then in place of Eq (\ref{result}) we will have $P_{\rm error} \propto N^{-(1+\alpha)/4}$.
\bibitem{careful_w/gauss}
Some care is required to be sure that the central limit behavior is actually valid in the regime of $z$ relevant for computing $P_{\rm error}$.  The key point is that $z_+$ grows more slowly than any power of $N$.
\bibitem{value}
The optimal error probability could be even larger if the cost of missing ordered behavior is greater than the cost of responding to spurious coincidences. 
\bibitem{KT_72}
D Kahenman \& A Tversky, {\em Cognitive Psych} {\bf 3,} 430--454 (1972).
\bibitem{TK_71}
A Tversky \& D Kahneman, {\em Psych Bull} {\bf 76,} 105--110 (1971).
\bibitem{sort_of}
Strictly speaking we can pose the problem only in terms of the decisions, but it seems plausible that at least an implicit estimate of the current bias will be a step in the process of decision making.  See \cite{sugrue+al_04} and forthcoming work from the same authors.
\end{thebibliography}
\end{document}